\documentclass[11pt, epsfig]{article}                              
\usepackage{epsfig}       
\usepackage{amssymb}     
\usepackage{amsfonts}

\newskip\humongous \humongous=0pt plus 1000pt minus 1000pt

\newif\ifdtup

\catcode`\@=11
 
\@addtoreset{equation}{section}
\def\theequation{\thesection.\arabic{equation}}
 
\def\@normalsize{\@setsize\normalsize{15pt}\xiipt\@xiipt
\abovedisplayskip 14pt plus3pt minus3pt%
\belowdisplayskip \abovedisplayskip
\abovedisplayshortskip \z@ plus3pt%
\belowdisplayshortskip 7pt plus3.5pt minus0pt}
 
\def\small{\@setsize\small{13.6pt}\xipt\@xipt
\abovedisplayskip 13pt plus3pt minus3pt%
\belowdisplayskip \abovedisplayskip
\abovedisplayshortskip \z@ plus3pt%
\belowdisplayshortskip 7pt plus3.5pt minus0pt
\def\@listi{\parsep 4.5pt plus 2pt minus 1pt
     \itemsep \parsep
     \topsep 9pt plus 3pt minus 3pt}}
 
\relax

\catcode`@=12
 
\catcode`\@=11
 
\def\section{\@startsection{section}{1}{\z@}{3.5ex plus 1ex minus
   .2ex}{2.3ex plus .2ex}{\large\bf}}
 
\def\thesection{\arabic{section}}    
\def\thesubsection{\arabic{section}.\arabic{subsection}}

\def\appendix{\setcounter{section}{0}
 \def\thesection{Appendix \Alph{section}}
 \def\thesubsection{\Alph{section}.\arabic{subsection}}
 \def\theequation{\Alph{section}.\arabic{equation}}}
\newcommand{\beq}{\begin{equation}}
\newcommand{\eeq}{\end{equation}}
\newcommand{\bea}{\begin{eqnarray}}
\newcommand{\eea}{\end{eqnarray}}
\newcommand{\beas}{\begin{eqnarray*}}
\newcommand{\eeas}{\end{eqnarray*}}

\newcommand{\non}{\nonumber}
\newcommand{\bquo}{\begin{quote}}
\newcommand{\enqu}{\end{quote}}

\def\Arg{\hbox {\rm Arg}}

\def\de{\partial}

\def\o{\over}
\def\im{\hbox{\rm Im}}

\def\Arg{\hbox {\rm Arg}}

\def\2{{1\over 2}}
\def\adt{{\dot\alpha}}

\def\alp{\alpha}

\def\tht{\theta}
\def\thb{{\bar \theta}}

\begin{document}

\begin{titlepage}
{\hfill     IFUP-TH/2003/44}  
\bigskip
\bigskip

\begin{center}
{\large  {\bf  
NON-ABELIAN SUPERCONDUCTORS - LESSONS FROM SUPERSYMMETRIC GAUGE THEORIES FOR QCD
 } }
\end{center}

\bigskip
\begin{center}
{\large  Kenichi KONISHI   
 }
\end{center}

\begin{center}
{\it   \footnotesize
Dipartimento di Fisica ``E. Fermi" -- Universit\`a di Pisa, \\
Istituto Nazionale di Fisica Nucleare -- Sezione di Pisa, \\
     Via Buonarroti, 2, Ed. C, 56127 Pisa,  Italy\\
  }

\end {center}

\noindent  
{\bf Abstract:}

  Much about the confinement  and dynamical symmetry breaking
 in QCD might be learned from
models with   supersymmetry.  In particular, models based on $N=2$ supersymmetric  theories 
with gauge groups $SU(N)$, $SO(N)$ and $USp(2 N)$    and with various number  of flavors, 
give  deep dynamical  hints about these phenomena.  
For instance, the  BPS non-abelian  monopoles   can  become the dominant degrees of
freedom in the infrared due to quantum effects.   Upon condensation  (which can be triggered in these class of
models by  perturbing them with  an adjoint scalar mass) they  induce confinement with calculable pattern of 
dynamical symmetry breaking.   This may occur either in a weakly interacting  regime or in a strongly coupled
regime (in the latter, often the low-energy degrees of freedom contain relatively non-local monopoles and dyons
simultaneously  and the system  is near a nontrivial fixed-point).  
  Also, the existence of  sytems with  BPS  {\it non-abelian vortices}  has been shown  recently.  These 
results   point toward 
 the idea that the ground state of QCD is a sort of dual  superconductor of non-abelian variety. 
\vfill  
 
\noindent   Talk given at the 
International Conference on Color Confinement and Hadrons in Quantum
                        Chromodynamics  ( ``Confinement 2003" ), Riken, Tokyo, July  2003

\begin{flushright}
November  2003
\end{flushright}  
\end{titlepage}

\section { Confinement in    $SU(N)$   YM   Theory \label{sec:dualsc} } 
  
The test charges in $SU(N)$   YM theory take values in   ($Z_N^{(M)}, Z_N^{(E)}$)  where $Z_N$  is the center of $SU(N)$ and  
$Z_N^{(M)}, Z_N^{(E)}$ refer to the magnetic and electric center charges. 
  ($Z_N^{(M)}, Z_N^{(E)}$)  classification of phases  follows \cite{tHooft,TH}.  (See Figures)  Namely, 

\begin{enumerate}  
 
\item    If field with  $x=(a, b)$ condense,   particles $X= (A,B)$ with  
$$    \langle x, X \rangle   \equiv     a \, B - b \, A \ne 0 \quad  (mod \,\, N)   $$
are confined.    (e.g.   $\langle \phi_{(0,1)}\rangle \ne 0$ $\to$   Higgs phase.)

\item  Quarks are confined if some field $ \chi$  exist, such that   
   $$\langle \chi_{(1,0)}\rangle \ne 0.$$  
\item  In the   softly broken $N=4$  (to $N=1$) theory   (often referred to as $N=1^*$)    all different types of massive vacua,
  related by $SL(2,Z)$, appear;  the  chiral condensates in each vacua are  known. 
\item 
{\it   Confinement index} \cite{CSW}     is   equal to  the smallest possible  $r \in Z_N^{(E)} $   for which  Wilson loop displays   no area law. 
For instance,  for   $SU(N)$  YM,      $r=N$   in the vacuum with complete confinement;     $r=1$  in the 
totally Higgs vacuum, etc.  

\item   In softly broken $N=2$  Gauge Theories,  dynamics turns out to be  particularly transparent.  

\end{enumerate}

\begin{figure}[ht]
\begin{center}
\leavevmode
\epsfxsize=6 cm      
\epsffile{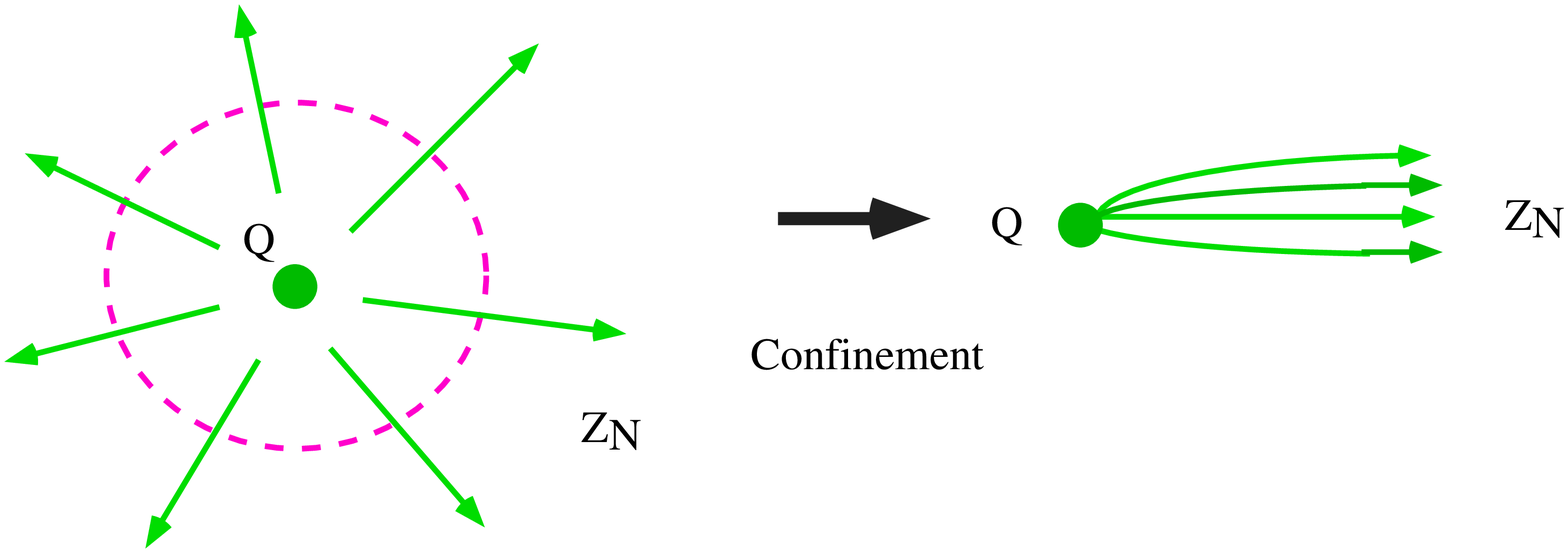}               
\end{center}
\end{figure}
    \begin{figure}[ht]
\begin{center}
\leavevmode
\epsfxsize=3  cm   
\epsffile{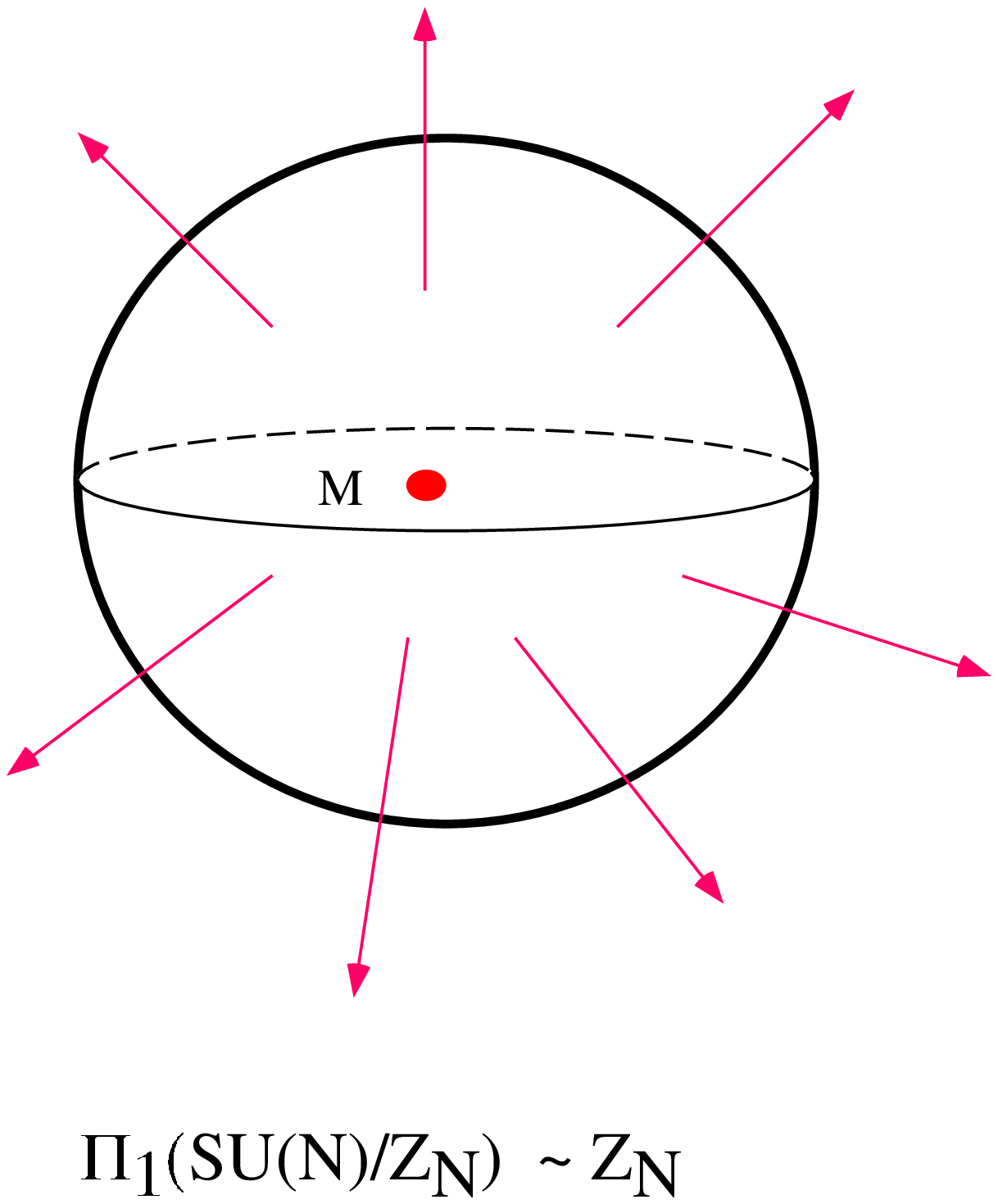}           
\end{center}  
\end{figure}

The questions we wish to address here are: 
 What is   $\chi $ in QCD?    How do they interact  ?   Is  chiral symmetry breaking  related to confinement?  $\theta$ vacua?;
  ${\epsilon^{'} \over \epsilon} $;  $\Delta I = {
1\over 2}$ ?

A familiar idea is that the ground state of QCD is a dual superconductor \cite{TH}.  
 There exist  no  elementary nor soliton  monopoles in QCD; however, 
monopoles  can be detected  as topological singularities (lines in $4D$)  of Abelian gauge fixing,  $SU(3) \to U(1)^2$.
Alternatively, one can assume that  certain  configurations  close  to the Wu-Yang monopole  
 ($SU(2)$)
  $$A_{\mu}^a=  {\tilde \sigma}(x) (
\partial_{\mu} { n}\times
  { n})^a  + \ldots, \qquad      { n}({ r})=   { {  r} \over r}  \quad  
\Rightarrow  \quad       A_i^a  = \epsilon_{aij}{r^j \over r^3}  $$
dominate \cite{FN}.
 Although there are some evidence  in lattice QCD   \cite{DG}
for ``Abelian dominance",  there are 
several questions to be answered.   
 Do abelian  monopoles carry flavor?   What is the form of ${ L}_{eff}$?
What about the  gauge dependence? 
Most significantly,     does dynamical  $SU(N) \to U(1)^{N-1}$ breaking occur?  That would imply \cite{MSYung} a
 richer spectrum of mesons  ($T_1 \ne T_2, $ etc)
not seen in Nature and not expected in QCD.  Both in Nature and presumably in QCD  
there is  only one ``meson" state
$$    {\hbox {\rm Meson} }  \sim\sum_{i=1}^N  |\, q_i \,  {\bar q}_i \, \rangle $$  
i.e.,  $1$  state  vs   $\left[{N \over 2}\right]$  states.  (See Figure.)   
Assuming   $SU(N) \to U(1)^{N-1}\times  Weyl$ symmetry  is not enough to solve the problem.   

\begin{figure}[ht]
\begin{center}
\leavevmode
\epsfxsize=2   cm 
\epsffile{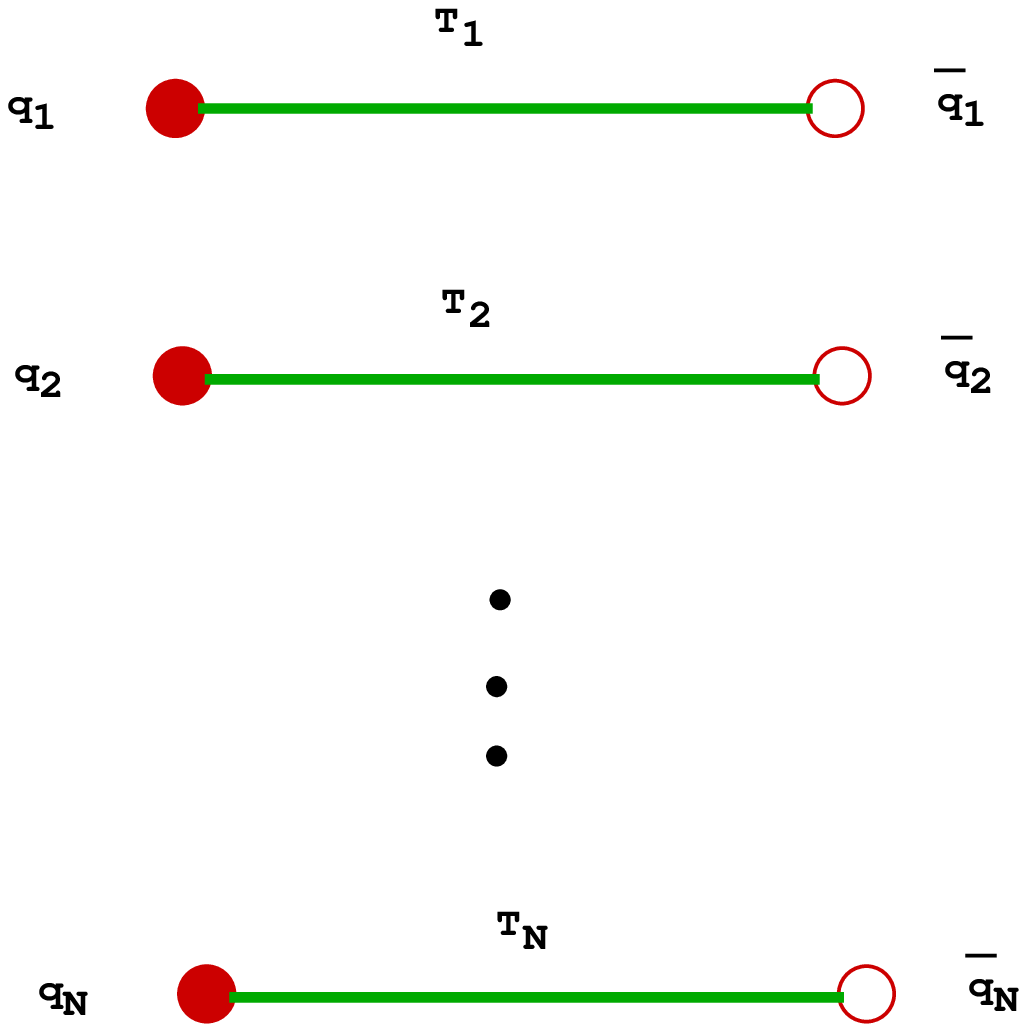}            
\end{center}
\end{figure}

In an attempt to answer these questions  we address ourselves to the world of supersymmetric gauge theories, 
where the non-abelian gauge dynamics can be analyzed to depth.  Supersymmetric  theories in fact continue to
 surprise us for rich insight they give us about the dynamics of 
non-abelian interactions. 
Although not a main subject of this talk, let me point out that much is going on at present in the 
exploration  of various $N=1$ or $N=2$ susy gauge theories \cite{CDSW}.

   Let us start by recalling some basic facts about supersymmetric gauge theories \cite{WB}. 

\section{SQCD}   

\subsection{ Basics of Susy gauge theories }

 The basic susy algebra contains   
$$  \{  Q_{\alp}, {\bar  Q}_{\adt} \} =  2 \sigma^{\mu}_{\alp, \adt}  P_{\mu}.  
$$  
   In order to construct supersymmetric theories it is convenient to introduce  superfields \cite{WB}   
$$   F(x, \tht, \thb) =  f(x) +  \tht \psi(x)   + \ldots   
$$
$$    Q_{\alp}= {\de \o \de \tht^{\alp} }  -  i \sigma^{\mu}_{\alp, \adt} \thb^{\adt} \de_{\mu},  \quad 
 {\bar  Q}_{\adt}=  {\de \o \de \thb^{\adt} }  -  i \tht^{\alp}   \sigma^{\mu}_{\alp, \adt} \de_{\mu},
$$
In general they are reducible with respect to  supersymmetry transformations. We construct smaller irreducible multiplets.   Chiral superfields
are defined by the constraint
$  {\bar D}
\Phi=0$   (
$  {D}
\Phi^{\dagger} =0$ )  so that  
$$     \Phi(x, \tht, \thb) =  \phi(y) + \sqrt2  \tht \, \psi(y) +   \tht \tht \,F(y), \quad   y=  x +  i \tht \sigma \thb 
$$
$$   D_{\alp}=  {\de \o \de \tht^{\alp} }  +  i \sigma^{\mu}_{\alp, \adt} \thb^{\adt} \de_{\mu}, \qquad {\bar  D}_{\adt}=  -   {\de \o \de
\thb^{\adt} }  -  i \tht^{\alp}   \sigma^{\mu}_{\alp, \adt} \de_{\mu},
$$
Verctor superfields are defined to be real 
   $V^{\dagger }= V.$  They are conveniently expressed in terms of a chiral (fermionic) superfield
$$      W_{\alpha} =  - { 1\o 4}   {\bar D}^2   e^{-V}   D_{\alpha}  e^V   
  = -i \lambda \, + \,{\mu,  
\o 2} \,  (\sigma^{\mu} \, {\bar \sigma}^{\nu})_{\alpha}^{\beta} \, F_{\mu \nu} \,
\theta_{\beta} + \, \ldots
$$
   Supersymmetric Lagrangian  ($\int d \tht_1 \,   \tht_1=1$, etc)    can then be written simply as 
\beq 
{\cal L}=      {1\over 8 \pi} \im \, \tau_{cl} \left[\int d^4 \theta \,
\Phi^{\dagger} e^V \Phi +\int d^2 \theta\,{1\o 2} W W\right]
+ \int    d^2 \theta\, W (\Phi) \label{Lagr}
\eeq
where   $ W (\Phi) $ is the   superpotential  and  
  $\tau_{cl} =  {  \tht \o  2 \pi} + { 4 \,  \pi \, i \o g^2}. $
The scalar potential is the sum of the $F$-term and $D$-term: 
$$  V_{sc} =    \sum_{mat} \left|  {\de W \o  \de \phi }  \right|^2  + { 1\o 2 } \sum_a \left|  \sum_{mat} \phi^* t^a \phi \right|^2.   
$$
  For SQCD,   $\{\Phi \} \to    Q  \sim  {\underline  N},    \,\,\,   {\tilde Q} \sim  {\underline  N^*}$  of $SU(N)$  
$$   G_F=   SU(n_f) \times  SU(n_f)  \times   U_V(1) \times  U_A(1) \times  U_{\lambda}(1). 
$$
The theory has a characteristic continuous vacuum degeneracy  - flat directions or classical moduli space  (CMS).  
 \begin{figure}[ht]
\begin{center}
\leavevmode
\epsfxsize=4cm  
\epsffile{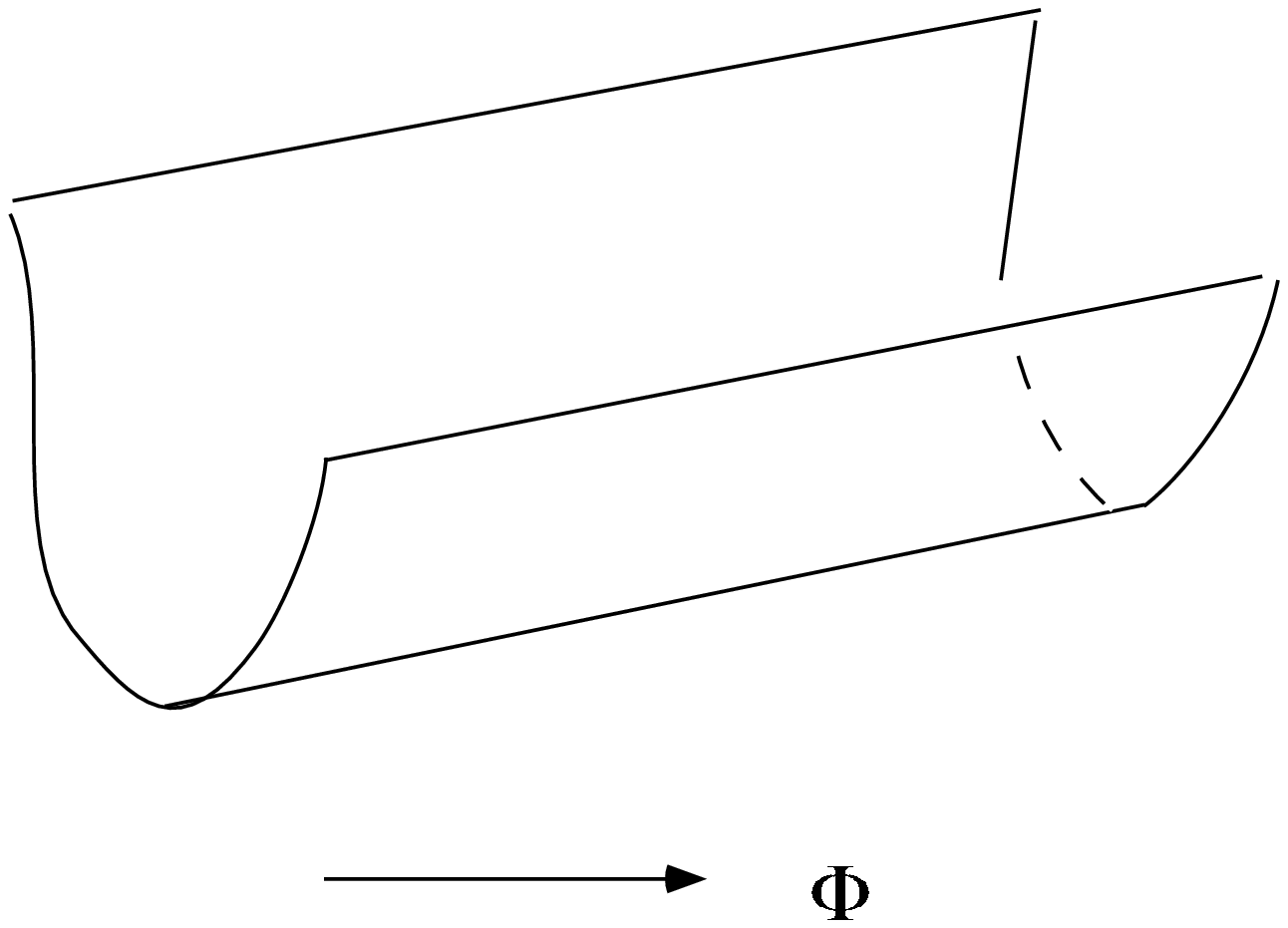}               
\end{center}
\label{susy}   
\end{figure}
For instance, it looks like  for $n_f < n_c$,     
$$    Q =  {\tilde Q}^{\dagger}    =  \pmatrix{a_1 & 0   & \ldots  &0     \cr
      0 &  \ddots  &    &   \cr 
  0 & \ldots  &   &   a_{n_f}    \cr
0 & 0 & \ldots  & 0  \cr
\ldots  &  &  & \ldots       }.
$$
   The problem is:   is superpotential generated dynamically?  Is   CMS   modified?

\subsection {  Phases of SQCD; Seiberg's  duality  }   
The analyses by use of various supersymmetry,   Ward-Takahashi identities, nonrenormalization theorems \cite{NSV},
symmetry considerations \cite{IS},
dynamical (instanton)  calculations  \cite{SV,AKMRV}  and Seiberg's  duality \cite{seib}   have established
the following picture of the vacua   in the massless SQCD: 

\begin{itemize}

\item  The dynamically generated  superpotential   implies  the  vacuum runaway  for  $n_f < n_c$,   
whereas   no  superpotential is generated  for  $n_f >   n_c. $

\item    For  $n_f  =  n_c$  the moduli space (space of vacua)  is quantum mechanically modified as 
$$  \det M  -  B \, {\tilde  B}  =\Lambda^{2 n_f}.   $$

\item   For  $ {3 n_c \o 2} < n_f <    3 n_c$ (conformal window), the system at the origin of the moduli space  is in an 
infrared fixed point  (superconformal theory - SCFT): the low energy phsyics is   described either as the original SQCD or
as  the  dual
$SU({\tilde n}_c)=  SU(n_f- n_c)$   theory with  dual quarks. 
   \begin{figure}[ht]
\begin{center}
\leavevmode
\epsfxsize=4  cm      
\epsffile{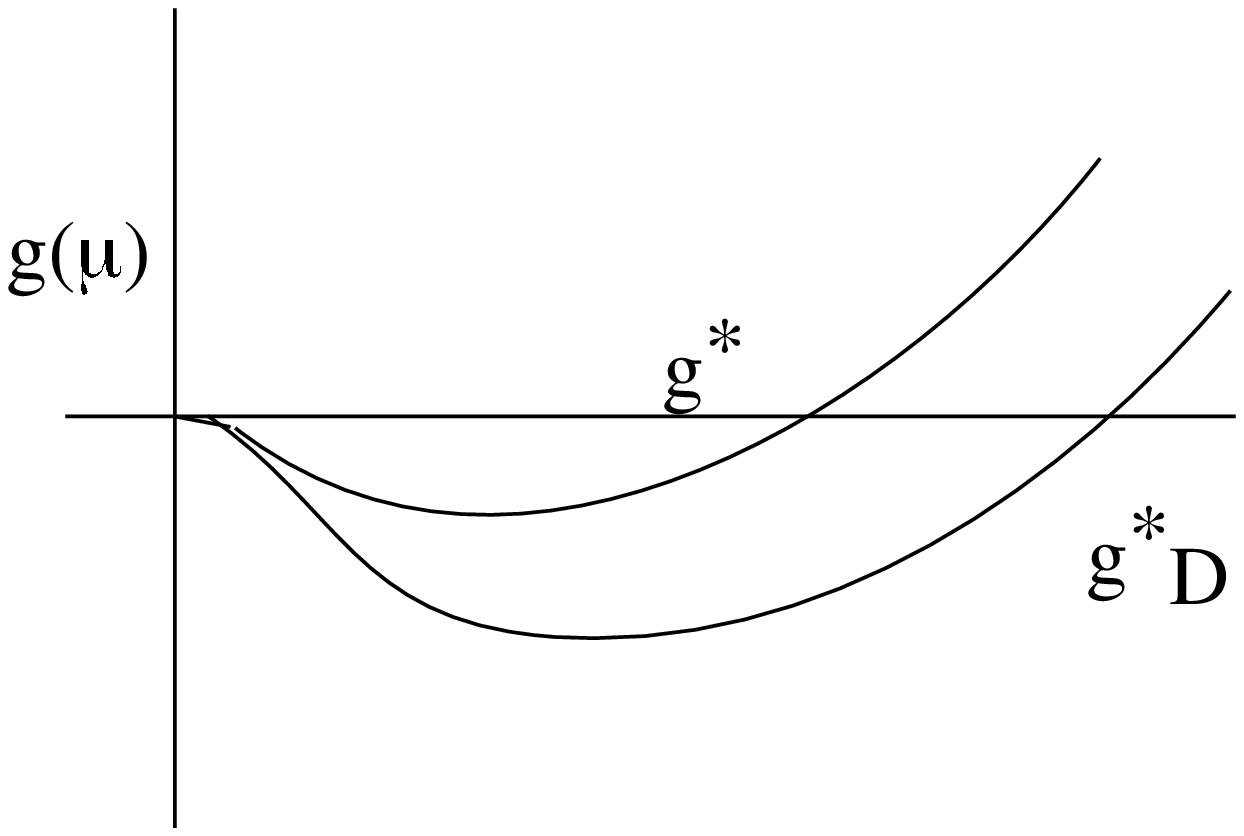}                
\end{center}\label{Seib}
\end{figure}
\end{itemize}    
\begin{center}
\footnotesize
\vskip .3cm
\begin{tabular}{|ccccc|}
\hline
 $N_f$    &   Deg.Freed.      &  Eff. Gauge  Group
&   Phase    &  Symmetry     \\
\hline
\hline
$0$ (SYM)   &   -    &   -              &   Confinement
   &     -           \\ \hline
$ 1 \le N_f  <    N_c $            &  -     & -       &
  no vacua        &    -      \\ \hline
$N_c$    &  $M, B, {\tilde B} $        &   -
  &    Confinement
&          $U(N_f)$ 
\\ \hline
$N_c +1$   &   $ M, B,{\tilde B} $       &   -
  &    Confinement
&         Unbroken   
\\ \hline
$N_c+1 < N_f < {3N_c \over 2} $    &   $ q, {\tilde q}, M  $     &   $ SU({\tilde N}_c ) $     &   Free-magnetic
&        Unbroken
\\ \hline
$ {3N_c \over 2} < N_f < 3 N_c  $   &   $q, {\tilde q},M  $   or   $Q, {\tilde Q} $    &   $SU({\tilde N}_c ) $  or  $ SU(N_c) $   &  SCFT
&        Unbroken  \\ \hline
$N_f  = 3N_c$   &   $Q, {\tilde Q} $   &   $SU(N_c ) $   &   SCFT (finite)
&        Unbroken  
\\ \hline
$N_f > 3N_c  $  &  $Q, {\tilde Q} $   &
$ SU(N_c)  $                &  Free Electric  
&    Unbroken     \\ \hline
\end{tabular}
\label{tabsun}
\end{center}

\section {Confining vacua in $N=2$ Supersymmetric Gauge Theories  }

Although dynamical properties of $N=1$ supersymmetric  gauge thoeries are thus  quite well known by now,  as in the
example of SQCD discussed above,  a detailed understanding of the working of confinement and dynamical symmmetry
breaking is still  lacking. 
$N=2$ theories   appear   to  allow for a  deeper level of  understanding of the nonperturbative dynamics, by displaying 
 the quantum behavior of magnetic monopoles and vortices  very clearly.   We start from afar: Dirac's monopoles in QED. 

\subsection {   Dirac's monopoles  }   

As is well known    QED admits  pointlike magnetic monopoles if   Dirac's quantization condition 
\begin{equation}   g \, e = { n \over 2}, \qquad   n \in   {\  Z},    \label{Dirac}
\end{equation}
is satisfied.  In the presence of a magnetic monopole,  there cannot be a gauge vector potential which is everywhere
regular. A possible singularity (Dirac string)  along   $ (0,0,0) \to (0,0,-\infty)$  is invisible if  (\ref{Dirac})   is satisfied.       A proper 
formulation is to  cover  $S^2$  by two  regions a ($0  \le    \theta < {\pi \over 2}+\epsilon$) and b (${\pi \over 2}- \epsilon < \theta \le 
\pi
$) 
\cite{WY} 
$$   (A_{\phi})^a =   { g \over r \sin \theta } (1 - \cos \theta), \qquad       (A_{\phi})^b =  -  { g \over r \sin \theta } (1 + \cos \theta),   $$
so that in each neighborhood the vector potential is regular.   The two descriptions are related along the equator  by  a
gauge transformation   
$$   A_i^a  =  A_i^b -U^{\dagger}   {i \over e   }\partial_i  U,  \qquad   U=  e^{2 i g e \phi}.    
$$    
The gauge transformation is well-defined if the condition (\ref{Dirac}) is  met. 
  More generally, for dyons $(e_1, g_1)$,   $(e_2 g_2)$,  the quantization condition  reads  
\begin{equation}   e_1 \, g_2 -  e_2  \, g_1 = { n \over 2}, \qquad   n \in   {\  Z}.       \label{Diracbis }  
\end{equation}
The topology involved is:         $\Pi_1(U(1))=  {\  Z}. $

\subsection {  Non-Abelian gauge theories }

In the case of a non-abelian gauge group,  one might embed Dirac's monopole in  a $U(1)$ subgroup.   
  However, the homotopy group properties such as 
\begin{equation}     SU(2) \sim S^3, \qquad      \Pi_1(SU(2))=  {\  1},
\end{equation}
\begin{equation}     SO(3) \sim {S^2 \over Z_2}, \qquad     \Pi_1(SO(3))=  {\  Z}_2,
\end{equation}
show  that  there are no topologically stable monopoles in $SU(2)$, $SU(N)$; there is only  one type of monopole in  $SO(3)$, and so on \cite{WY}. 

In spontaneously broken gauge theories,  instead, there are  ('t Hooft-Polyakov)  monopoles \cite{TP}  
$$   SU(2)    \,\,\,{\stackrel {\langle \phi \rangle    \ne 0} {\longrightarrow}}     \,\,\, U(1)    $$  
 $$   {D}  \phi    \,\,\,{\stackrel {r \to   \infty  } {\longrightarrow}}   \,\,\,0    ,   \quad    \Rightarrow   \quad 
\phi \sim   U \cdot  \langle \phi \rangle  \cdot U^{-1};  
$$ 
$$   A_i \sim  U \cdot {\partial_i  }  U^{\dagger}  \quad  \Rightarrow    \quad  F_{ij}=   \epsilon_{ijk }  { r_k   \over r^3}    m  {\tau_3
\over 2}  
$$
($m=\pm1, \pm2, \ldots$)   which are regular, finite energy soliton-like configurations,  with 
   topology,          
     $ \Pi_2( SU(2)   /U(1)   )  =
\Pi_1(U(1) )=  {\  Z}. $
 The static energy can be written as    (Bogomolny)  
$$   H= \int d^3x  \left[    { 1\over 4}  (F^a_{ij} )^2   +  { 1\over 2}  (  D_i \phi^a )^2  + {\lambda \over 2}  (\phi^2 - v^2)^2 \right].  
$$
$$  =  \int d^3x  \left[     { 1\over 4}  (F^a_{ij}  - \epsilon_{ijk} D_k \phi^a )^2 +  { 1\over 2}  F^a_{ij} \epsilon_{ijk} D_k \phi^a  + {\hbox {\rm pot.}} 
\right]  
$$
where   $ { 1\over 2}  F^a_{ij} \epsilon_{ijk} D_k \phi^a =  \partial_i S_i; 
\,\, \,   S_i=   { 1\over 2} \epsilon_{ijk}   F^a_{ij} \phi^a:$   the second term in the square bracket is a topological invariant.  It follows
that  in a given sector
 $$H \ge   \int d^3x  \nabla\cdot { S} =  {4 \pi v \over g}  m,  \qquad  \,\,\,m=1,2,\ldots  $$
If  $\lambda=0$  the  configuration of the minimum energy is given by the solution of the linear  (Bogomolny) equations
$$  F^a_{ij}  - \epsilon_{ijk} D_k \phi^a=0; \qquad B^a_i =   D_i \phi^a
$$
whose  solutions are known in analytic form.    

  A more general situation is that of a  spontaneously broken  gauge theory
 $$   G   \,\,\,{\stackrel {\langle \phi \rangle    \ne 0} {\longrightarrow}}     \,\,\, H  $$  
where $H$ is non-abelian \cite{GNO,EW}.  The asymptotic behavior is 
 $$   {D} \phi    \,\,\,{\stackrel {r \to   \infty  } {\longrightarrow}}   \,\,\,0    ,   \quad    \Rightarrow   \quad 
\phi \sim   U \cdot  \langle \phi \rangle  \cdot U^{-1}  \sim       \Pi_2(G/H)  =
\Pi_1(H);
$$ 
$$   A_i \sim  U \cdot {\partial_i  }  U^{\dagger}   \to     \quad  F_{ij}=   \epsilon_{ijk }  { r_k   \over r^3}    \beta_{\ell}   T_{\ell}.
$$
where  $   T_i  \in       {\hbox {\rm  Cartan S.A.  of }}   \,  H.   $  
Topological quantization leads to  $2 \, {\ \alpha }  \cdot {\ \beta  } \subset {\  Z}$, where 
$   \beta_i  =   {\hbox {\rm  weight vectors  of}} \,\,  {\tilde H} =  {\hbox {\rm  dual   of}} \,\, H.   $
Examples of the dual of several  groups  are given in the Table below. 

\begin{table}[h]     
\caption{  ${\tilde H}     \Leftrightarrow   H  $}  
\begin{center}
\begin{tabular}{c  c   c}
\hline  
$SU(N)/Z_N       $        &   $\Leftrightarrow$                 &    $SU(N)     $          \\
  $ SO(2N)  $     &   $\Leftrightarrow$    &   $SO(2N) $       \\  
  $ SO(2N+1)  $     &   $\Leftrightarrow$     &   $USp(2N) $       \\ \hline
\end{tabular}
\end{center}
\end{table}

The point of this discussion  is that these objects -- abelian and non-abelian BPS   monopoels  --
appear naturally in  $N=2$ gauge theories as dynamical degrees of freedom  
and play a crucial role in the infrared physics.    Before coming to this
central issue, however,  let us first
briefly review the celebrated Seiberg-Witten solution of
$N=2$ gauge theories \cite{SW1,SW2,curves}.  

\subsection {  Seiberg-Witten, $N=2$ Gauge Theories   }

 \begin{figure}[ht]
\begin{center}
 \leavevmode
  \epsfxsize=6 cm   
\epsffile{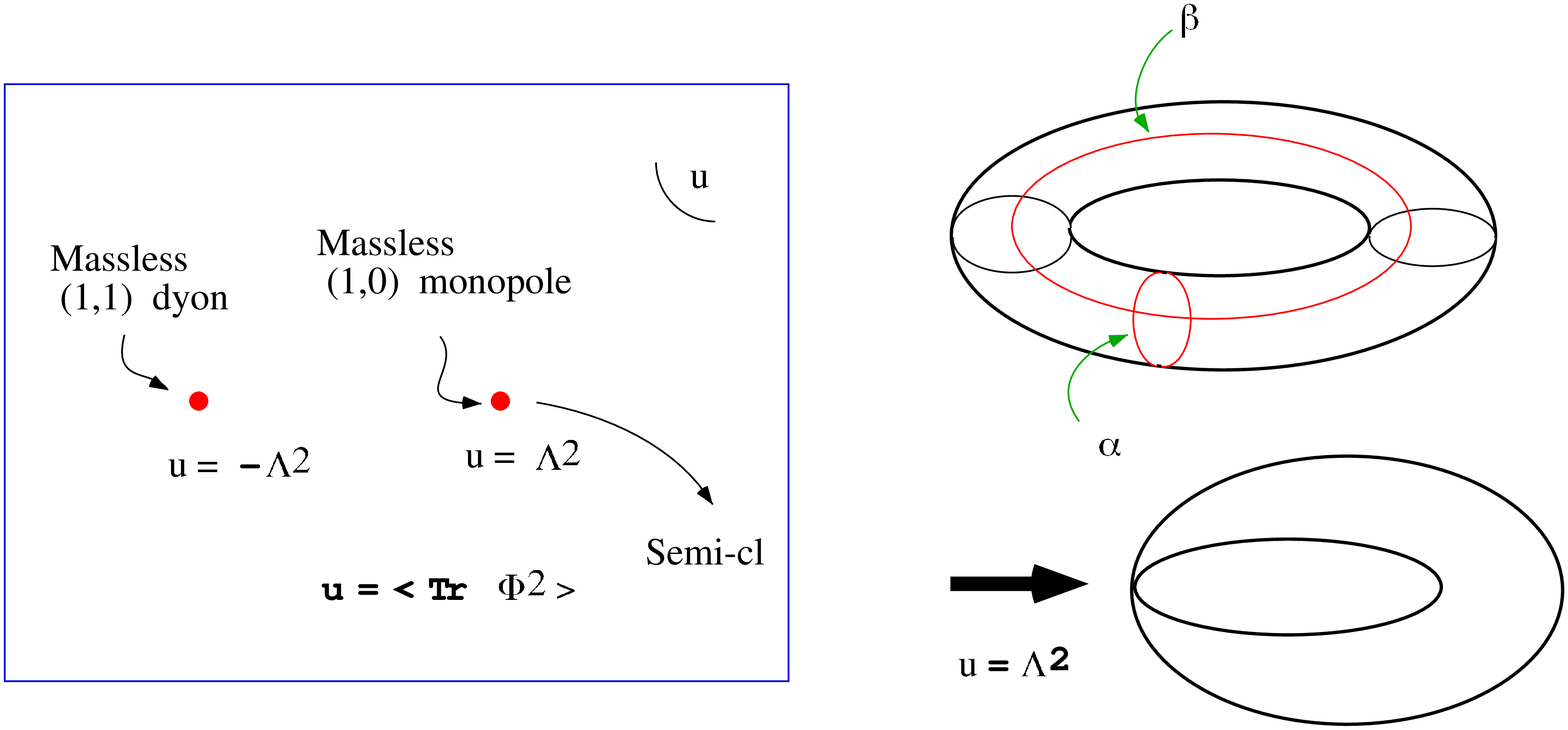}           
\end{center}   
\end{figure}    
         
The      Lagrangian of a $N=2$   YM theory is  Eq.(\ref{Lagr})   with $W (\Phi) =0,$  where $\Phi$  is a chiral superfield in the adjoint
representation of the gauge group.     For
$SU(2)$  
 the vacuum degeneracy
(moduli space) is parametrized  as  
\begin{equation}   \langle \Phi \rangle    =        \pmatrix { a  & 0  \cr   0 & -a},   
\end{equation}  
 $a \ne 0$   breaks   the gauge symmetry as  $SU(2)  \to U(1)$.  In the infrared,   
$$  L_{eff}  =  {\hbox {\rm Im}} \left[    \int  d^4\tht  \,  {\bar A}  \,\, { \partial F_p(A) \over \partial A}    +  \int { 1 \over 2}  
  { \partial^2  F_p(A) \over \partial  A^2  } W_{\alp}
W^{\alp} \right]  
$$
where   $W_{\alp},   A$  describe $N=2$,    $U(1)$ theory.      $ F_p(A)$  is called  prepotential. 
 Define   the dual of   $A$,  $A_D  \equiv   { \partial F_p(A) \over \partial  A}$: then     
$$  {d \, A_D \over d \, u} =  \oint_{\alpha} { dx \over y},  \qquad   {d \, A \over d \, u} =  \oint_{\beta} { dx \over y},  $$   
where the curve    ($u \equiv  {\hbox {\rm Tr}} \, \langle  \Phi^2  \rangle $  describes the quantum moduli space - QMS)  is 
 $$  y^2= (x-u) (x+ \Lambda^2) (x  -  \Lambda^2).   $$
The exact mass formula   (BPS) following from the   $N=2$  susy algebra is
$$    m_{n_m, n_e}  = \sqrt 2\,  | \,  n_m  A_D  +  n_e   A  \, |.      
$$
The  above four formulae consitute the  Seiberg-Witten  solution for the pure $N=2$ Yang-Mills theory \cite{SW1}. The
solution has been extended to more general gauge theories 
\cite{SW2,curves}.        

The adjoint scalar mass ( $\mu \, \Phi^2$  perturbation ) leads to the low-energy  effective  superpotential near the singularity, $u 
\simeq \Lambda^2$: 
$$ W_{eff}  =    \sqrt {2}   \, A_{D}  \,   M  {\tilde M}  +  \mu \,  U (A_{D}) 
$$
Minimization of the potential leads to the condensation of the monopole  $\langle M \rangle \sim \sqrt{\mu \, 
\Lambda }\,\,$   (Confinement).

It is interesting to note that   at the singularities $u= \pm \Lambda^2$,  instanton sum diverges
$$    \langle  {\hbox {\rm Tr}} \Phi^2   \rangle =   {a^2 \over 2} +  {\Lambda^4  \over a^2 } + \ldots  =  \ldots +  1 + 1 + 1 + \ldots  
$$

The discussion can be generalized to $N=2$  pure YM theories with other gauge groups.     In general,   dynamical abelianization occurs near the 
monopole singularities,  for instance, $SU(N)$ gauge group gets dynamically broken as     
$SU(N)
\to U(1)^{N-1}$  ({\it cfr}  QCD).    

It is important to realize that    these light ``monopoles" are indeed 't Hooft-Polyakov monopoles becoming light by quantum 
corrections.   This can be proven by studying the  charge fractionalization \cite{KT,Fer}.  For instance, the electric charge 
of the monopole  is known to behave as  
 $$
{2 \over g} Q_e =  n_e +
\left[ -{4 \over \pi}\Arg\, a + {1 \over 2 \pi}
\sum_{f=1}^{N_f} \Arg \, (m_f^2 -2 a^2)\, \right] \, n_m + \ldots 
\label{chargesc} 
$$  
in the semi-classical region where $a$ is the adjoint scalar VEV.   The Seiberg-Witten exact solution, when
extrapolated back to the semi-classical domain, reproduces  exactly this results.
An analogous check has been done for the quark-number fractionalization \cite{KT}.  
 There is an interesting phenomenon of quantum quenching of quark numbers  of massless, condensing monopole.  
Also, the non-abelian flavor quantum numbers of the  monopoles as encoded in the Seiberg-Witten solution are  consistent with 
the well-known  Jackiw-Rebbi mechanism.

 \subsection { More general $N=2$    models  }

The study of the more general class of $N=2$ theories \cite{curves,APS,CKM}   has shown that  there are variety of
confining vacua (see Fig. ~\ref{QCDvac}):

\begin{figure}[ht]
\begin{center}
\leavevmode
\epsfxsize=6    cm 
\epsffile{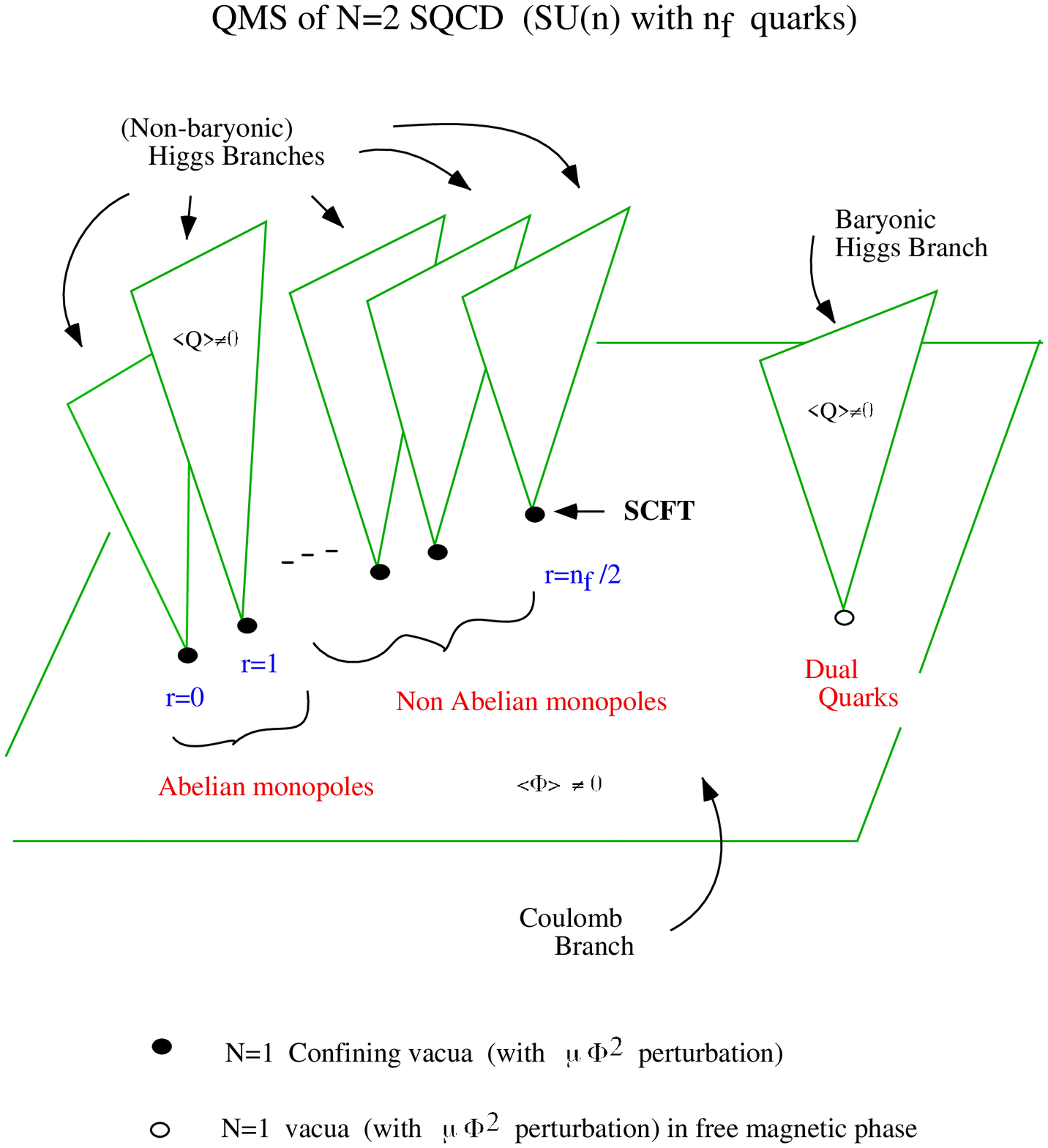}           
\end{center}
\label{QCDvac}
\end{figure}
   
\begin{enumerate}  

\item There are  vacua ($r=0,1$)  in which the low-energy effective action is an abelian (dual) gauge theory. 
Upon the adjoint scalar mass perturbation $\mu \, \Phi^2$, the magnetic monopole condenses (confinement);  the system displays 
dynamical abelianization, a feature not shared by the real world QCD; 
\item In a series of $r$-vacua,  the effective  action is   $G_{eff} \sim SU(r) \times U(1)^{n_c-r-1}$;   with  $n_f$ dual quarks     in
${\underline r}$ of the low-energy $SU(r)$ group.  The ``dual   quarks" can be identified with the standard non-abelian monopoles
\cite{BK}. 
  These  $r$-vacua  exist for     $r\le {n_f \over 2}$;    
\item   Superconformal theory  (SCFT) occurs   at $r={n_f \over 2}.$
Here the question is what (mutually nonlocal) degrees of freedom describe the SCFT (which confines upon the perturbation $\mu \, \Phi^2$). 

\end{enumerate}

Physics of    $ USp(2\, n_c) $   (  $ SO(n_c) $ similar ) theory is also very  interesting.   
 All $r$ vacua  at finite $m$  are identical to those in the   $SU(n_c)$  theory.     They   collapse into a single   SCFT
at $m \to 0$; 
  in other words,   all confining vacua  of the theory with a vanishing bare quark masses  (with perturbation  $\mu \,
\Phi^2$) are deformed SCFT, with mutually nonlocal  dyons in   the infrared \cite{CKM}. 
Also,    the global symmetry breaking pattern   ( $SO(2\, n_f) \to   U(n_f)$  in the case of  $ USp(2\, n_c) $   theory ) in
these deformed  SCFT  is very reminiscent of what happens in QCD  ({\it cfr}    $ \langle {\bar \psi } \, \psi \rangle ^{(QCD)} 
\ne 0$).

\begin{figure}[ht]
\begin{center}
\leavevmode
\epsfxsize= 6   cm 
\epsffile{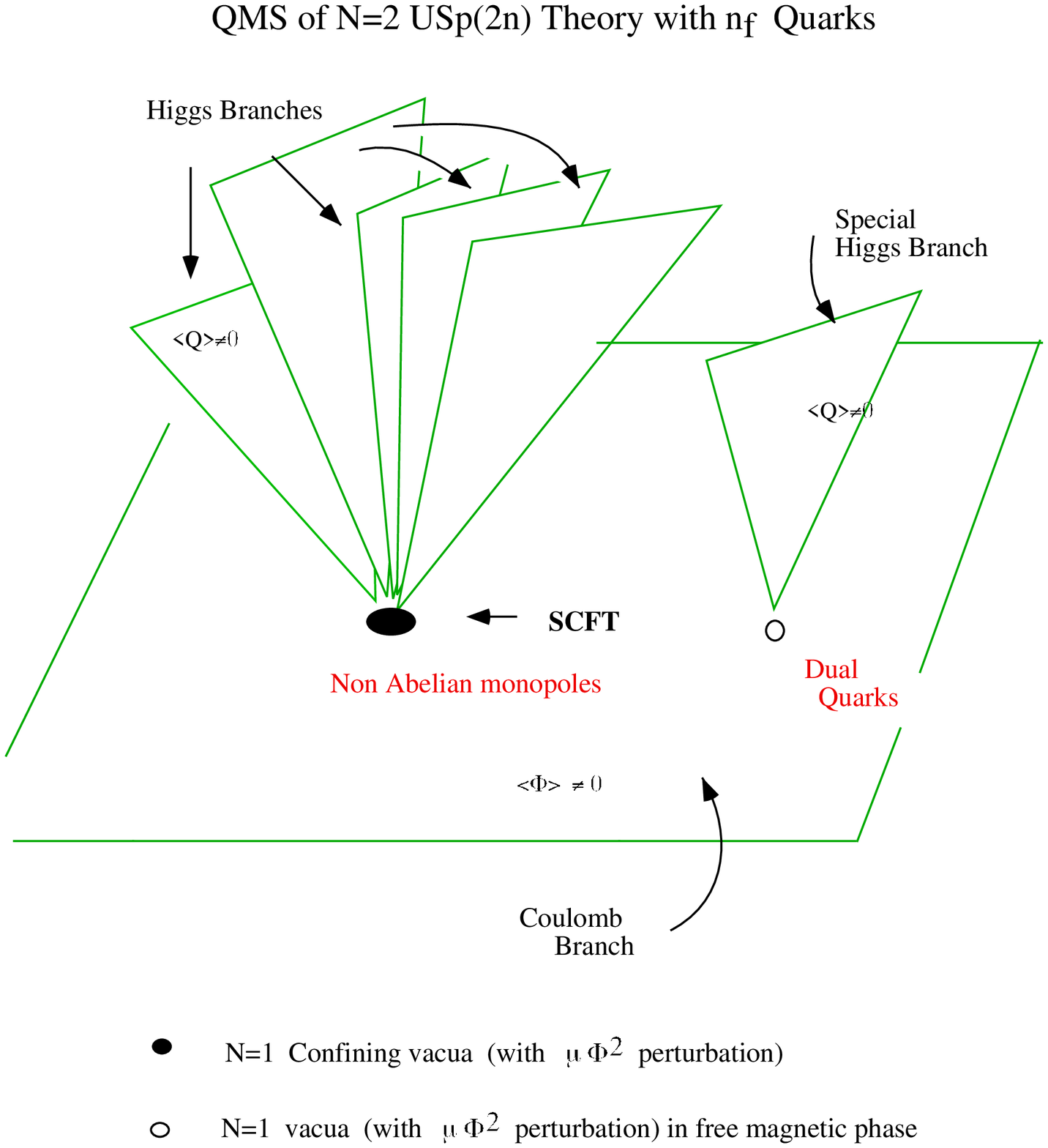}             
\end{center}
\end{figure}

\subsection {  More about  non-abelian  Monopoles } 

Consider the system with gauge symmetry breaking 
$$     SU(3) {\stackrel {\langle \phi \rangle } {\longrightarrow}}     SU(2) \times  U(1), \qquad   \langle \phi \rangle = 
 \pmatrix{  v & 0& 0  \cr  0 & v & 0 \cr  0&0& -2v }
$$
 By making use of the   't Hooft-Polyakov solutions   in $SU_U(2),  SU_V(2)     \subset SU(3) $  
one finds   two degenerate $SU(3)$ solutions.  
\begin{table}[h]      
\begin{center}
\begin{tabular}{c  c   c}   
  monopoles    &    ${\tilde  {SU}  (2)  } $           &    ${\tilde U(1) }     $          \\    \hline 
  ${\tilde q}  $        &     $ {\underline 2 }$     &      $1$       \\  \hline 
\end{tabular}
\end{center}  
\end{table}   
Analogously, for  the system with symmetry breaking  
$$     SU(n) {\stackrel {\langle \phi \rangle } {\longrightarrow}}     SU(r) \times  U^{n-r}(1), \qquad   
\langle   \phi\rangle = 
 \pmatrix{  v_1  { 1}_{r\times r}   & { 0 } &  \ldots &   { 0}   \cr  { 0 }  & v_2   & 0   & \ldots   \cr  { 0}  &0& \ddots 
&  
\ldots
\cr { 0}    & 0   &  \ldots  & v_{n-r+1}   }   
$$
one finds the following set of the minimal  monopoles:

\begin{table}[h]      
\begin{center}
\begin{tabular}{c  c   c c c  c c  }
  monopoles    &    ${\tilde  {SU}}  (r)   $           &    ${\tilde U_0(1) }     $      &    ${\tilde U_1(1) }     $     &    ${\tilde U_2(1) }  $   &
$\ldots  $  &         ${\tilde U_{ n-r-1} (1) }  $        
\\    \hline 
  $ q  $        &     $ {\underline r   }$     &      $1$     &  $0$ &   $0$ & \ldots   &  $0$    \\  \hline
 ${ e}_1  $        &     $ {\underline 1 }$     &     
$0 $     &   $1$  &  $0$     & $\ldots $     &  $0$    \\  \hline 
${ e}_2 $        &     $ {\underline 1  }$     &   $0 $     & $0 $  &  $1$   & $0$    &  $0$  \\  \hline 
$\vdots $        &     $ {\underline 1   }$     &      $0 $     &  $ \ldots $    &   &   &   $0$    \\  \hline 
${ e}_{n-r-1}    $        &     $ {\underline 1  }$     &      $0$     & $0$     &  $ \ldots$   &  $\ldots $     &  $1$   \\  \hline 
\end{tabular}
\end{center}
\end{table}  
We note \cite{BK}   that 
    they represent a degenerate  $r$-plet of monopole solutions   ($ q$);  
that  they have  {\it       the same charge structure as that of the ``dual quarks" appearing  in the  $\, r\,$-vacua of  $ N=2$  SQCD}. The latter
 have also the   correct flavor quantum numbers  as expected from  the  Jackiw-Rebbi    
mechanism.

\subsection {  Subtle are non-abelian monopoles    \label{sec:non-abelian}}    

There are      certain subtleties about the non-abelian monopoles. 
First of all,  ``colored dyons"  are known not  to  exist \cite{CDyons}.    More precisely, 
 in the background of a 't Hooft-Polyaknow monpole it is not possible to define a globally defined set of 
generators  isomorphic to those of   $H$.     This being so,   it
does not preclude the non-abelian monopoles of our interest:  magnetic particles having  abelian and non-abelian  charges, {\it both} magnetic, can
perfectly well exist, and do appear in the $r-$ vacua of the sofly broken $N=2$ SQCD \cite{BK}.     
   The main point is that the non-abelian monopoles  are multiplets of the dual ${\tilde H}$  group, not of $H$ itself. Transformations among the
members of a multiplet of monopoles are nonlocal transformations with respect to the original gauge field variables. 

 The no-go theorem  nonetheless  means that the true gauge symmetry of the theory is not  
$  H \otimes   {\tilde H},  
$
as  sometimes suggested, but    $ {\tilde H},$  $H$  or some other combination, according to which physical degrees of
freedom are   relevant in a particular situation.

Also,  it  is not  justified to study the system
$   G   \,\,\,{\stackrel {\langle \phi \rangle    \ne 0} {\longrightarrow}}     \,\,\, H  $    
as a limit of maximally broken cases as sometimes made in the  literature.  
 non-abelian monopoles  are never really  semi-classical, even if    $$\langle \phi \rangle  \gg \Lambda_H.$$    
For  $H$ were  broken  it would produce simply an  {\it  approximately }  degenerate  set of  monopoles
as, for instance,  in  the  pure $N=2$,    $SU(3)$ theory.  
Only  if  $H$ remains   unbroken  do   non-abelian monopoles in  irreducible representation   of  ${\tilde H}$  appear.  Which option is realized is a
dynamical question which cannot be determined from a semiclassical consideration. 

 The important fact is that the second option {\it is}  realized in the 
  $r$ vacua of $N=2$  SQCD  with 
   $SU(r)  \times   U(1)^{n_c-r+1} $  gauge group, where $r < { N_f \over 2}$.       
This last constraint  can be understood from a renormalization-group consideration: for   $r < { N_f \over 2}$      
there is a sign flip in the  beta  functions of the dual magnetic gauge group, with respect to that in the underlying theory:     
\[     b_0^{(dual)} \propto   -  2 \, r  +   n_f  >  0,   
\qquad      b_{0} \propto  -  2 \, n_c +    n_f  <     0.     
\label{betafund}   \]
 In fact, when such a  sign flip  is  not possible,  {\it e.g.}, pure $N=2$ YM,    
dynamical abelianization occurs!  
  {\it The quantum behavior of  non-abelian monopoles  thus depends  critically on the presence of  massless fermions in the underlying 
theory.}  
 $r= {n_f \over 2}$  is a boundary case: the corresponding   vacua  are   SCFT  (nontrivial  IR  fixed point).
Non-abelian monopoles and dyons   still  show up as low-energy degrees of freedom,   but their interactions are
nonlocal and strong. The possible mechanism of confinement in these vacua has been recently studied \cite{AGK}.

\subsection {${ \bf  Z}_N $    Vortices } 

Once the relevant degrees of freedom which act as the order parameter of confinement are  identified,  we are  interested in the dominant field
configurations which are capable of actually confining   quarks.  
In the abelian dual superconductor picture of confinement in a $SU(N)$   YM theory, 
the quarks would be   confined by abelian Abrikosov-Nielsen-Olesen vortices of $U(1)^{N-1}$.  
However, this leads to the difficulty mentioned at the end of the Section \ref{sec:dualsc}.
The quarks must be confined by some sort of non-abelian chromoelectric vortices. 

The simplest type of vortices involving a non-abelian gauge group is the 
$ {\bf  Z}_N   $    vortex, which    occurs in a system with gauge symmetry breaking 
    $$  SU(N)  \Rightarrow {\bf Z}_N.    $$   An analogous vortex appear in a system with a general symmetry breaking pattern,  $H   \Rightarrow 
{ C}
$,   a discrete center. Vortices represent nontrivial elements of 
$\Pi_1( H /  { C})$,    e.g.      $\Pi_1(SU(N) /  {\bf Z}_N   ) =  { \bf  Z}_N.$  
 The asymptotic behavior of the fields is
$$   A_i  \sim   { i \over g}    \, U(\phi) \partial_i  U^{\dagger}(\phi);    
\quad     \phi_A \sim   U \phi_{A}^{(0)}  U^{\dagger},   \qquad  U(\phi) = \exp{i \sum_j^r 
\beta_j T_j
\phi}
$$
where $T_j$  are the generators of the Cartan subalgebra of $H$.  The quantization condition is   
($\alpha$ =  root vectors of $H$) 
$$  U(2 \pi)  \in    {\bf Z}_N, 
\qquad   
{ \alpha} \cdot {  \beta}  \in  {\bf Z}:        
$$  
the vortices are characterized by  the {\it weight vectors} of  the group ${\tilde H}$, dual of $H$.   
It seems as though  the vortex solutions were classified according to the irreducible representations   of   ${\tilde   H }  =  SU(N)$.
Actually, the fact  that the topology involved is $\Pi_1(SU(N) /  {\bf Z}_N   ) =  { \bf  Z}_N$  means that the stable vortices are 
characterized by    ${\bf Z}_N$ charge   ($N$-ality)  only \cite{KS}.    

These  ${\bf Z}_N$  vortices are non BPS and this makes the analysis of these objects   so far relatively little explored.   However there are
interesting  quantities which characterize these systems such as the 
 tension ratios  for different $N$-ality sources:  an  intriguing  proposal  (sine formula) \cite{DS,Sine}   is  
$$  T_k  \propto     \sin { \pi \, k \over N}       $$ 
which can be measured on the lattice.

\subsection{       BPS     vortices;   non-abelian Superconductors     }

Systems with BPS  vortices with a non-abelian flux - non-abelian superconductors -  have been shown   to exist only recently
\cite{ABEKY}\footnote{An apparently  similar, but different model   has been  studied by
Hanany and Tong \cite{HT}. }.  Consider a gauge theory  in which the gauge group is broken at two very different scales
    $$     G   \,\,\,{\stackrel {\langle \phi \rangle
     \ne 0} {\longrightarrow}}     \,\,\, H   \,\,\,{\stackrel {\langle \phi^{\prime}  \rangle
     \ne 0} {\longrightarrow}}   \,\,   \emptyset, \qquad   \langle \phi \rangle \gg      \langle \phi^{\prime}  \rangle, 
$$
where   the  unbroken (non-Abelian) group $H$   gets  broken completely  at   a much lower scale, $ \langle \phi^{\prime}  \rangle$. 
We are interested in the physics at scales between the two scales $ \langle \phi \rangle $ and $ \langle \phi^{\prime}  \rangle $. 
When $\Pi_1(H) \ne \emptyset$ the system develops vortices.   
   If the theory contains an  exact continuous symmetry  $G_F$, respected both by the interactions and by the vacuum
(not spontaneously broken),   but broken by a vortex solution, then  there will be  a nontrivial degeneracy of vortex solutions
(zero modes). 

An  example \cite{ABEKY}  is the   $SU(3)$  $N=2$ theory  with $n_f =4, 5 $
 quark flavors   with large common (bare) mass $m,$  with the $N=2$ symmetry broken softly to $N=1$  by the adjoint mass term,
$\mu \, {\hbox {\rm Tr}} \, \Phi^2$.   We consider a particular vacuum, the ``$r=2$"  vacuum,   of   this system, which is characterized by the VEVs    
$$
\Phi = -{1\over \sqrt{2}}\left(
\begin{array}{ccc}
  m & 0 & 0 \\
  0 & m & 0 \\
  0 & 0 & -2m
\end{array}  \right),  \quad  <q^{kA}>=<\bar{\tilde{q}}^{kA}>=\sqrt{\frac{\xi}{2}}\left(
\begin{array}{cc}
  1 & 0  \\
  0 & 1  \\
  \end{array}\right)    
$$
where   $\xi = {\mu \, m}$.    We take the (bare) quark mass $m$ much larger than  $\mu$  so that   $m \gg \sqrt{\xi}$. 
At the mass scales between $m$  and   $\sqrt{\xi}$,  the system has an exact $SU(2) \times U(1)/ Z_2 $ gauge symmetry
as well as an $SU(n_f)$  global symmetry. The action has the form, after 
the Ansatz $\Phi= \langle \Phi \rangle $;    $ q = {\tilde q}^{\dagger}$;   and $ q \to { 1\over 2}  q $:  
$$
S=\int d^4x \left[\frac1{4g^2_2} \left(F^{a}_{\mu\nu}\right)^2 +
\frac1{4g^2_1}\left(F^{8}_{\mu\nu}\right)^2
+ \left|\nabla_{\mu}
q^{A}\right|^2\right.
$$
\begin{equation}
\left.
+ \frac{g^2_2}{8}\left(\bar{q}_A\tau^a q^A\right)^2+
\frac{g^2_1}{24}\left(\bar{q}_A q^A - 2\xi
 \right)^2  \right]. 
\label{len}
\end{equation}
The tension can be rewritten \`a la Bogomolny: 
\begin{eqnarray}  
T   &=&\int{d}^2 x   \Big(    \sum_{a=1}^3  \left[\frac1{2g_2}F^{(a)}_{ij } \pm
     \frac{g_2}{4}
\Big(\bar{q}_A\tau^a q^A  )  
\epsilon_{ij} \right]^2    +  \nonumber\\  
& + & 
\left[\frac1{2g_1}F^{(8)}_{ij} \pm 
     \frac{g_1 }{4\sqrt{3}}
\left(|q^A|^2-2\xi \right)
\epsilon_{ij }\right]^2   + \frac{1}{2} \left|\nabla_i \,q^A + i  \,\varepsilon \, \epsilon_{ij}
\nabla_j\, q^A\right|^2
\pm
\frac{\xi}{2  \sqrt{3}}\tilde{F}^{(8)}       \Big)    \non 
\end{eqnarray} 
where the first three terms are positive definite and    the fourth term is a topologically  invariant,
$U(1)$ flux. 
 The non-abelian Bogomolny equations 
$$ \frac1{2g_2}F^{(a)}_{ij } \pm
     \frac{g_2}{4}
\Big(\bar{q}_A\tau^a q^A  )  
\epsilon_{ij} =0, \qquad     (a=1,2,3);
$$  $$     \frac1{2g_1}F^{(8)}_{ij} \pm
     \frac{g_1 }{4\sqrt{3}}
\left(|q^A|^2-2\xi \right)
\epsilon_{ij }=0, 
$$
\begin{equation} 
\nabla_i \,q^A +i \varepsilon\epsilon_{ij}
\nabla_j\, q^A =0, \qquad   A=1,2.
\label{naB}\end{equation}
follow  from the last formula. 
The equations  (\ref{naB})    have abelian  ($n,k$)   solutions of the type (where  $n,k$ are integers)   studied in \cite{MY}     
$$
q^{kA}=\left(
\begin{array}{cc}
  e^{ i \, n\,\varphi  }\phi_1(r) & 0  \\
  0 &  e^{i \, k \,  \varphi }\phi_2(r) \\
  \end{array}\right),
$$
$$
A^3_{i}(x) = -\varepsilon\epsilon_{ij}\,\frac{x_j}{r^2}\
\left((n-k)-f_3(r)\right),\;
$$
\begin{equation}
\label{sol}
A^{8}_{i}(x) = -\sqrt{3}\ \varepsilon\epsilon_{ij}\,\frac{x_j}{r^2}\
\left((n+k)-f_8(r)\right)\,
\end{equation}
 where  $\phi_1(r),$  $\phi_2(r),$  $f_3(r)
$,   $f_8(r)$  are profile functions  with appropriate boundary consitions. 

The crucial observation is that the system (\ref{len})  has an exact 
   $SU(2)_{C+F} $   symmetry,  which  is neither broken  by the interactions nor  by  the squark  VEVS.  
However,  an individual vortex configuration breaks it 
 as   $ SU(2)_{C+F}  \to U(1)$   therefore the vortex acquires  zero modes  parametrizing 
       $$  { SU(2) \over U(1) }  \sim  { CP}^1 \sim S^2.  $$     
For instance,   minimum  vortices of generic orientation (all degenerate) 
 can be explicitly constructed as $$  
q^{kA}=U\left(
\begin{array}{cc}
 e^{  i \varphi}   \phi_1(r) & 0  \\
  0 &  \phi_2(r) \\
  \end{array}\right)U^{-1}=   e^{\frac{i}{2}\varphi  (1+n^a\tau^a)} \,  U\left(
\begin{array}{cc}
  \phi_1(r) & 0  \\
  0 &  \phi_2(r) \\
  \end{array}\right)U^{-1},
$$
$$
{ A}_{i}(x) = U  [- {\tau^3\over 2} \, \epsilon_{ij}\,\frac{x_j}{r^2}\,
[1-f_3(r)] ]   U^{-1} = -\frac12\,n^a \tau^a\epsilon_{ij}\,\frac{x_j}{r^2}\,
[1-f_3(r)],
$$
\begin{equation}
\label{rna}
A^{8}_{i}(x) = -\sqrt{3}\ \epsilon_{ij}\,\frac{x_j}{r^2}\,
[1-f_8(r)],
\end{equation}
where     
$  U   $  is an  $SU(2)$ matrix, which smoothly interpolate between  the abelian  $(1,0)$  and $(0,1)$ vortices. 
Explicitly,  if  $n^a=(\sin \alpha \cos \beta,  \sin \alpha  \sin \beta , \cos \alpha)$, the rotation matrix is given
by   $U = \exp{ -i \beta\, {\tau_3 /  2} } \cdot   \exp{ -i \alpha  \, {\tau_2 / 2} } $.

The (massive)  non-abelian monopoles  resulting from the gauge symmetry  breaking 
$SU(3)  \to SU(2) \times U(1)/ Z_2$ by the adjoint $\Phi$ VEV,   are confined by these non-abelian monopoles.

~~~
\begin{figure}[ht]
\begin{center}
\leavevmode
\epsfxsize 5  cm
\epsffile{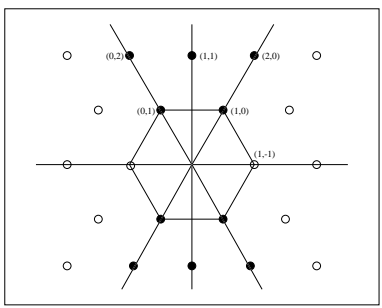}           
\end{center}
\label{abelian}
\caption{The spectrum of the abelian  vortices  in the  $U(1) \times U(1)$  gauge theory in the Higgs phase
\cite{MY}.  }
\end{figure}

\begin{figure}[ht]
\begin{center}
\leavevmode
\epsfxsize 5  cm
\epsffile{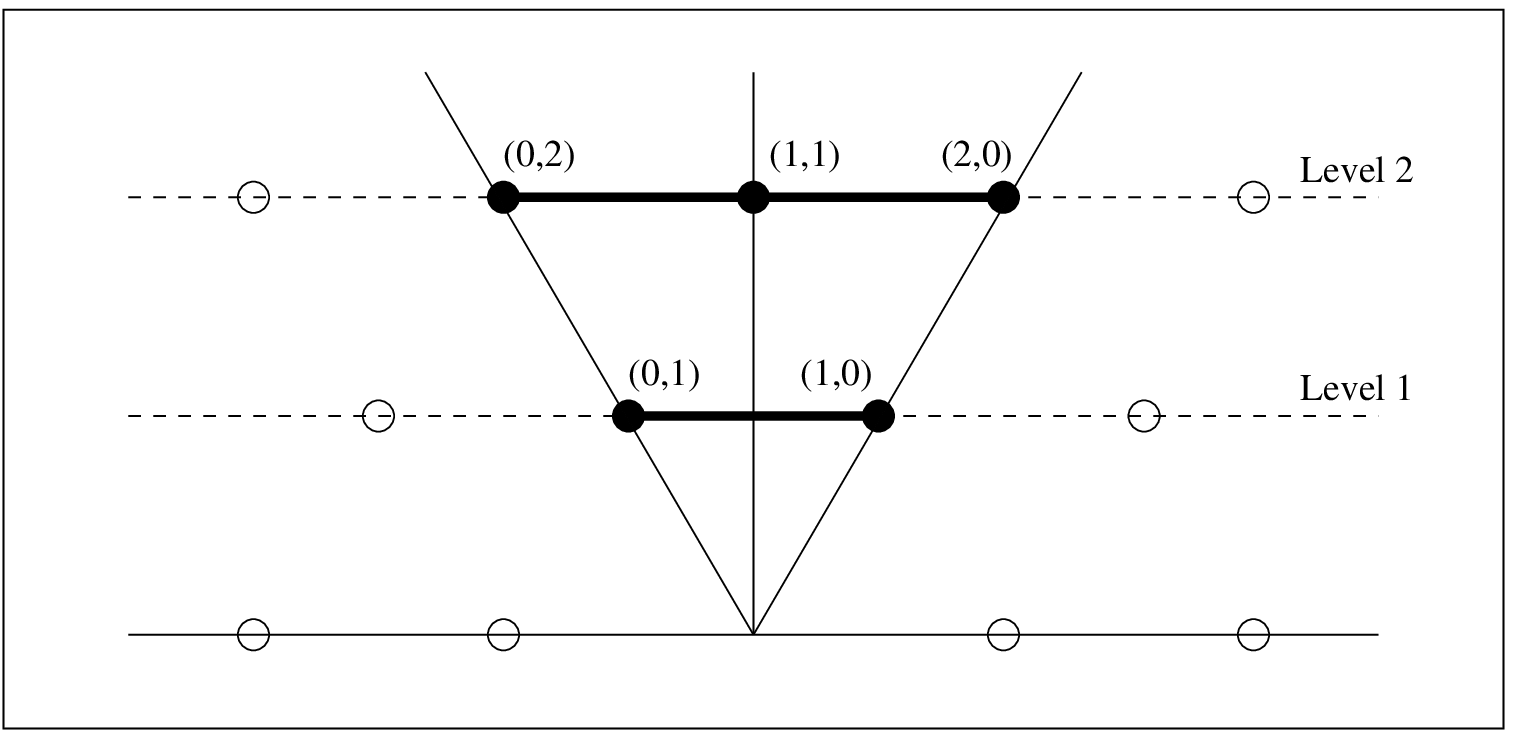}           
\end{center}
\label{non-abelian}
\caption{The reduced spectrum of the non-abelian  vortices  in the  $SU(2) \times U(1)/{\mathbb Z}_2 $  gauge
theory  (Eq.(\ref{len}))   in the Higgs phase
\cite{ABEKY}.  }
\end{figure}

\subsection  {\bf Subtle are (also)  non-abelian vortices   \label{ABEKY}: }

 The  reduction of    the vortex spectrum   - meson spectrum   (see Figures above) -  is  due to the 
topology change 
$$
\Pi_1( \frac{U(1)\times U(1)}{{\bf Z}_2} )= {\bf Z}^{2} 
\quad  \to   \quad  
\Pi_1( \frac{SU(2)\times U(1)}{{\bf  Z}_2} )= {\bf   Z};    
$$ 
which occurs in the limit of equal masses  $m_i\to m$.  
  The transition from the  abelian  ($m_i \ne m_j$)   to the  non-abelian ($m_i=m)$   superconductivity is here reliably
and quantum mechanically     analyzed   as  the $SU(2) \times  U(1)$  subgroup  is non asymptotically free for  $n_f=4$ or $n_f=5$.  
 {\it  Note that the quantum behavior of the non-abelian vortices also depends crucially on the massless fermions present in 
 the underlying theory \cite{ABEKY}.}
For instance, in the $N=2$, $SU(3)$ theory with $n_f$ less or equal to $3$,  there are no quantum vacua with  unbroken $SU(2)$  gauge group.

Existence of degenerate, non-abelian vortices which continuously interpolate from $(1,0)$ to $(0,1)$ vortices, imply a corresponding  $SU(2)$
doublet of  monopoles transforming  continuously between them, as the latter act as
the sources (or sink)  of the former  (see Figure below)  when the full $SU(3)$  interactions are taken into account.

  The dynamics of vortex zero modes can be shown to be equivalent to the two-dimensional $O(3)={\bf CP}^1$ sigma model
($  {\bf n}  \to    {\bf n}(z,t)   $): 
$$ 
S^{(1+1)}_{\sigma}=\beta \int dz \, dt   \,   \frac12 \left(\partial \, n^a\right)^2  + {\hbox {\rm fermions}} .   
$$
It is dual \cite{VH,NSVZPR} to  a chiral  theory     
with  two vacua.  The exact $SU(2)_{C+F}  $  symmetry is not spontaneously broken.  The   
dual  ($N=1$) $SU(2)$  theory  is in confinement phase  and  has, correctly,  two vacua (Witten index).  

The whole picture generalizes naturally to  the sytem with the symmetry breaking    \beq SU(N)\to  {SU(N-1)  \times   U(1) \over {\bf  Z}_{N-1} }
 \to  \emptyset\,\,\,\eeq     with
$2 N  >    N_f \ge   2 (N-1)  $  flavors. 
The system at intermediate scales  has vortices    with  $ 2 (N-2)$-parameter family of   zero modes  representing  
$$ {SU(N-1)  \over SU(N-2) \times    U(1)} \sim  {\bf CP}^{N-2}. $$ 

   The analysis of \cite{ABEKY}  was made at large $m$  (large $\langle \phi \rangle$)  where the system is semiclassical.  
It is the existence of this large disparity of scales   ($m \gg \sqrt{\xi}$)   which allows us to treat approximately 
both  the BPS monopoles and vortices   as  stable configurations. This and other  questions  on
non-abelian monopoles and vortices   will be expounded  in a forthcoming paper \cite{ABEKM}.

 Though more difficult to
analyze, the situation at small $m$  where  the non-abelian monopoles condense and the quarks are confined by non-abelian chromoelectric vortices, 
is related smoothly  to  the non-abelian superconductor studied here,  via holomorphic dependence of the physics on $m$ and though 
the isomonodromy (in which quarks become monopoles and vice versa).

{~~}  
\begin{figure}[ht]
\begin{center}
\leavevmode
\epsfxsize 5  cm
\epsffile{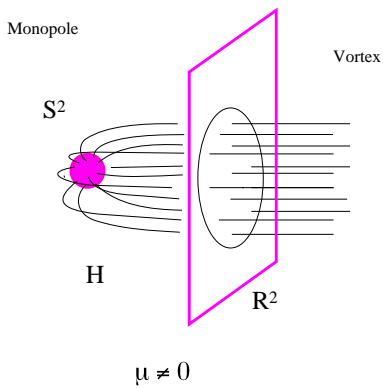}      
\end{center} 
\caption{A minimum non-abelian vortex ends on a minimum non-abelian monopole with an exactly matched
flux
\cite{ABEKM}. }
\end{figure}

\subsection {Lessons from $N=2$ SQCD} 

Summarizing, softly broken $N=2,$   $SU(n_c)$ gauge theories with  $n_f$  quarks with $m=0$, show different types of 
confining vacua:  

  \begin{enumerate}    
\item  $r =0,1$ vacua   are  described  by weakly coupled abelian monopoles;  

\item  $r < {n_f  \over 2}:$  The ground state is a non-abelian superconductor;  non-abelian monopoles condense and
confines the quarks; 

\item   $ r={n_f  \over 2}$ is a boundary case: the ground state is a deformed  SCFT,   with strongly  coupled non-abelian monopoles and dyons.  
 
\end{enumerate}  

In the $USp(2n_c)$ and $SO(n_c)$   gauge theories all confining vacua are   of the third type. 

\noindent    Both at  generic  $r$ - vacua  and  at the SCFT ($r={n_f \over 2}$) vacua of   $SU(n_c)$ SQCD, non-abelian monopoles condense
as
$$ \langle    { M}_{\alpha}^i  \rangle  = \delta_{\alpha}^i \, v \ne 0, \qquad  (\alpha=1,2, \ldots, r; \quad    i=1,2, \ldots, n_f)     $$
(``Color-Flavor-Locked  phase").   There are some indications that a similar result holds in the  $ r={n_f  \over 2},$
 almost conformal vacua \cite{AGK}.

\section {Hints for  QCD} 

What can one learn from these studies for QCD?   First of all, 
 dynamical abelianization  neither  is  observed in the real world  nor  is believed  to occur in QCD. 
On the other hand,    QCD  with $n_f$  flavors    and its possible dual  have the beta function coefficients (${\tilde n}_c =2, 3$,  $n_f=2,3$)  
such that
$$  b_0=   11 \, n_c - 2 \,n_f \quad  vs     \quad   {\tilde b}_0  =   11 \, {\tilde n}_c -  n_f,  $$
where we assumed  that in the standard QCD, the flavor carrying monopoles are scalars. Because of the large coefficient (eleven) in front of the
color multiplicity,  a 
sign  flip  (weakly-coupled non-abelian  monopoles) is hardly possible, even though in nonsupersymmetric theories higher loop contributions are
also important.   These two facts, together,  leave the option    of a strongly-interacting non-abelian superconductor, as in the almost
superconformal vacua in the
$N=2$  gauge theories, as the most likely picture of the ground state of  QCD. 
 
  Taking a more detailed hint from  supersymmetric models \cite{CKM} one might assume that  non-abelian
magnetic monopoles of QCD  condense   in a color-flavor-diagonal form
$$  \langle     { M}_{L, \alpha}^i  \rangle  = \delta_{\alpha}^i \, v_R \ne 0, \quad 
\langle     { M}_{R, i}^{\alpha}   \rangle  = \delta_i^{\alpha} \, v_L \ne 0,   
$$ 
 $(\alpha=1,2,\ldots {\tilde n}_c;   i=1,2,\ldots n_f).$ 
 As they are strongly coupled, a better physical picture   might be 
$$   \langle    { M}_{L, \alpha}^i  \, { M}_{R, j }^{\alpha}  \rangle  =  {\hbox {\rm const.}} \, \delta^i_j\ne 0;  $$
which  yields   for $ {\tilde n}_c=2, $   $n_f=2$ the  correct symmetry breaking pattern 
$$   G_F = SU_L(2)\times SU_R(2)  \Rightarrow   SU_V(2),  $$   
observed in Nature.   $U_V(1)$   (baryon number)  is not broken spontaneously in the real world,  therefore  the massless 
non-abelian monopoles  would have to possess  a vanishing   quark number.   
Intriguingly,  precisely such a phenomenon occurs  
in supersymmetric theories as a result of     nonperturbative renormalization \cite{KT}.

 \end{document}